\documentclass{elsart}

\usepackage{graphics}
\usepackage{epsfig}
\usepackage{amsfonts}
\usepackage{amssymb}
\usepackage{cite}
\usepackage{here}

\newcommand{\pd}{\phantom{0}}   
\newcommand{\pp}{\phantom{.}}   
\newcommand{\pff}{\phantom{.0}}  

\newcommand{\dd}{\mbox{\rm d}}

\begin{document}
\begin{frontmatter}

\title{Energy dependence of forward $^{1\!}S_0$ diproton production in the
$pp\to pp\pi^0$ reaction}

\author[dubna]{V.~Kurbatov},
\author[ikp]{M.~B\"uscher},
\author[dubna]{S.~Dymov},
\author[dubna]{D.~Gusev},
\author[ikp]{M.~Hartmann},
\author[erlangen,tbilisi]{A.~Kacharava},
\author[munster]{A.~Khoukaz},
\author[dubna]{V.~Komarov},
\author[dubna]{A.~Kulikov},
\author[dubna,tbilisi]{G.~Macharashvili},
\author[munster]{T.~Mersmann},
\author[dubna,ikp]{S.~Merzliakov},
\author[gatchina]{S.~Mikirtytchiants},
\author[ikp]{D.~Prasuhn},
\author[ikp]{F.~Rathmann},
\author[ikp]{R.~Schleichert},
\author[ikp]{H.~Str\"oher},
\author[dubna]{D.~Tsirkov},
\author[dubna]{Yu.~Uzikov},
\author[london]{C.~Wilkin\corauthref{cor1}}
\ead{cw@hep.ucl.ac.uk} \corauth[cor1]{Corresponding author.},
\author[erlangen]{S.~Yaschenko}.

\address[dubna]{Laboratory of Nuclear Problems, Joint Institute for Nuclear
  Research, 141980 Dubna, Russia}
\address[ikp]{Institut f\"ur Kernphysik, Forschungszentrum J\"ulich, 52425
  J\"ulich, Germany}
\address[erlangen]{Physikalisches Institut II, Universit\"at
Erlangen--N\"urnberg, 91058 Erlangen, Germany}
\address[tbilisi]{High Energy Physics Institute, Tbilisi State University, 0186
Tbilisi, Georgia}
\address[munster]{Institut f\"ur Kernphysik, Universit\"at M\"unster,
48149 M\"unster, Germany}
\address[gatchina]{St. Petersburg Nuclear Physics Institute, 188350 Gatchina,
  Russia}
\address[london]{Physics and Astronomy Department, UCL, London, WC1E 6BT, UK}

\begin{abstract}
The $pp\to\{pp\}_s\pi^0$ differential cross section has been
measured with the ANKE spectrometer at COSY--J\"ulich for seven
proton beam energies $T_p$ between 0.5 and 1.97\,GeV. By selecting
proton pairs with an excitation energy of less than 3\,MeV it is
ensured that the final $\{pp\}_s$ system is in the $^{1\!}S_0$
state. In the measured region of $\theta_{pp}^{\rm cm} \lesssim
18^\circ$, the data reveal a forward dip for $T_p\leq 1.4\,$GeV
whereas a forward peaking is seen at 1.97\,GeV. The energy
dependence of the forward cross section shows a broad peak in the
0.6--0.8\,GeV region, probably associated with $\Delta(1232)$
excitation, and a minimum at 1.4\,GeV. Some of these features are
similar to those observed for the spin--isospin partner reaction,
$pp\to d\pi^+$. However, the ratio of the forward differential
cross sections of the two reactions shows a significant
suppression of single pion production associated with a
spin--singlet final nucleon pair.
\end{abstract}

\end{frontmatter}
%
%
\newpage

Single pion production in nucleon--nucleon collisions,
$NN\rightarrow NN\pi$, is one of the principal tools used in the
investigation of $NN$ dynamics at intermediate
energies~\cite{GarcilazoMizutani,MachnerJP,HanhartPR}. Because of
the large momentum transfers involved, even close to threshold,
such a meson production process is sensitive to the
short--distance part of the $NN$ interaction.

The $pp\to d\pi^+$ reaction has been the subject of extensive
experimental study with the measurement of many spin observables,
as well as of the differential cross section from threshold up to
several GeV. However, the information that this provides is
restricted to final $NN$ states with spin $S=1$ and isospin $T=0$.
On the other hand, the $pp\to \{pp\}_s\pi^0$ process is
kinematically very similar to this provided that the excitation
energy in the final proton pair is very small. In this case, due
the the Pauli principle, the protons must be in the singlet
$^{1\!}S_0$ state, \textit{i.e.} have quantum numbers
$(J^P,\,T,\,L) = (0^+,1,0)$ compared to the
$(1^+,0,0\,\textrm{and}\,2)$ of the deuteron. Pion production in
the two cases therefore involves different transition matrix
elements so that a combined study of the two processes should
yield greater insight into the reaction dynamics.

If we consider the $pp\to d\pi^+$ reaction as the limit of triplet
$\{pn\}_t$ production in the $pp\to\{pn\}_t\pi^+$ channel, where
the strong final state interaction produces the
deuteron~\cite{migdal-watson}, the ratio of $\pi^0$ to $\pi^+$
cross sections will provide information on the relative strength of
spin--singlet to spin--triplet production, \textit{i.e.}\ give
information on the relative probability of pion production at
short distances in channels with different spin orientation of the
final nucleons. Because of the smallness of the signal, attempts
to identify spin--singlet production directly from data on the
$pp\to pn\pi^+$ cross section have only yielded upper
limits~\cite{UzikovWilkinPL,AbaevPL,Abdel-BaryPL}.

A small value of the singlet--triplet ratio is expected near
threshold since $s$--wave isovector pion rescattering is absent
for $pp\to pp\pi^0$ and heavy ($\omega$) meson exchange provides
the largest driving term~\cite{HanhartPL}. A small value of the
ratio is also predicted for energies around 0.4--0.6\,GeV since
the $S$--wave $\Delta(1232)N$ intermediate state that dominates
the $pp\to d\pi^+$ cross section is forbidden by conservation laws
in the $pp\to \{pp\}_s\pi^0$ case. The theoretical situation at
higher energies is largely open. The position is rather similar on
the experimental side since, away from the low energy domain, the
only other published data in the $^{1\!}S_0$ conditions were
limited to energies $T_p \leq 0.425\,$GeV~\cite{BilgerNP}.

We have previously reported a measurement of the
$pp\to\{pp\}_s\pi^0$ differential cross section obtained using the
ANKE spectrometer~\cite{BarsovNIM} for the single beam energy of
0.8\,GeV~\cite{DymovPL} . Here the two protons were detected at
small angles with respect to the incident beam and cuts were made
such that the excitation energy $E_{pp}$ in the final $pp$ system
was less than 3\,MeV. Under these conditions we expect the final
$\{pp\}_s$ pair to be almost purely in the singlet $^{1\!}S_0$
state. It was found for this beam energy that the angular
variation was rather strong but that at all the measured angles
the cross section was orders of magnitude smaller than that for
$pp\to d\pi^+$. To study the energy dependence of these effects,
we present here further measurements taken over a wide range of
energies, 0.508, 0.625, 0.700, 1.100, 1.400, and 1.970\,GeV.

The magnetic spectrometer ANKE is placed at an internal beam
station of the COSY cooler synchrotron of the Forschungszentrum
J\"ulich. Fast charged particles, resulting from the interaction
of the proton beam with the hydrogen cluster--jet
target~\cite{KhoukazEPJ} and passing through the analysing
magnetic field, were recorded in the forward detector (FD)
system~\cite{DymovPNL}. The FD system includes multiwire
proportional chambers for tracking and a scintillation counter
hodoscope for energy loss and timing measurements. The triggers
employed required the crossing of the two planes of the hodoscope
by at least one charged particle (SP--trigger) or by two particles
(DP--trigger)~\cite{DymovJC}. When the DP--trigger was used for
data taking, the SP--trigger ran (prescaled) in parallel for
luminosity measurement and calibration purposes. The tracking
system provided momentum resolution $\sigma_p/p \approx 1\%$ in
the range of interest and resolution in the excitation energy of
$\sigma(E_{pp}) \approx 0.2-0.8\,$MeV for events with $E_{pp} <
3\,$MeV.

Additional details of the experimental setup and the measurement
procedure are to be found in
refs.~\cite{DymovPL,DymovPNL,KomarovPL,YaschenkoPRL}. The data at
0.8\,GeV and above were taken during a single beam--time run,
whereas the lower energy results were obtained from other ANKE
calibration runs.

From measurements of the momenta of two charged particles in ANKE,
the $pp\to pp\pi^0$ channel was isolated by determining the
missing mass $M_x$ in the reaction. In more than 80\% of cases
where two fast particles were detected, the tracks passed through
different counters of the forward hodoscope. For these events the
particles could be clearly identified as protons on the basis of
the timing information. The difference of the arrival times of the
two particles measured by the counters was compared with the
time-of-flight difference deduced from the measured momenta
assuming that the particles both had the mass of the proton. For
the remaining $\approx 20\%$ of events, the hypothesis was made
that the two particles were indeed protons. As already shown for
the 0.8\,GeV data~\cite{DymovPL}, the missing--mass distributions
for both classes of events are very similar with only a slightly
enhanced background when no timing information was available. The
two sets were therefore combined in the subsequent analysis.

\begin{figure}[htb]
\begin{center}
\epsfig{figure=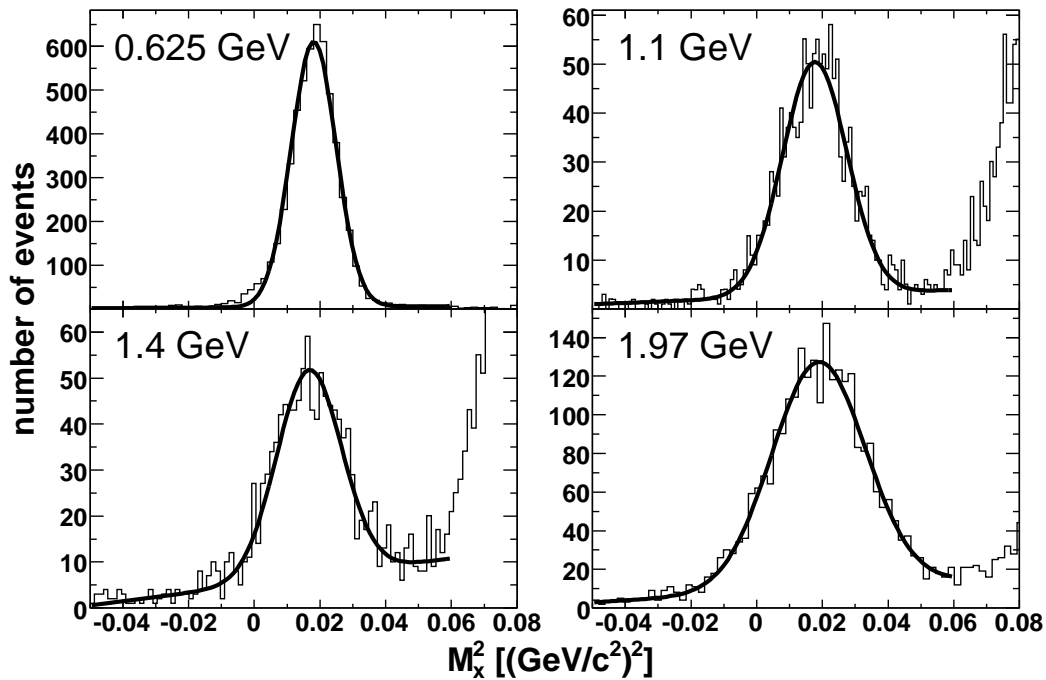,width=1\textwidth}
\caption{Missing--mass--squared distributions of the $pp\to ppX$
reaction for $M_x$ in the $\pi^{0}$ region. The curves show the
fits to the experimental spectra in terms of a Gaussian plus a
straight line. The data near $M_x^2=0$ were excluded from the fit
since, as indicated here by the 0.625\,GeV results, there may be
some $pp\to \{pp\}_s\gamma$ events in this region.}
\label{MX2_fit}
\end{center}
\end{figure}

Figure~\ref{MX2_fit} presents examples of the measured
missing--mass distributions in the $\pi^0$ region for events with
$E_{pp}<3$\,MeV. In addition to the $\pi^0$ peak, a rise of counts
is seen on the right hand side due to two--pion production which
seems to be largest at 1.1 and 1.4\,GeV. This background gives
little contribution in the $\pi^0$ region and the data were fitted
as a sum of a Gaussian and straight line. The region close to
$M_x^2=0$ was excluded from the fit since, as discussed for the
0.8\,GeV data~\cite{DymovPL}, there is the possibility here of
some contribution from the $pp\to \{pp\}_s\gamma$ reaction. In all
cases the peak position was consistent with $m_{\pi^0}$ to within
the experimental uncertainties of about $\pm10\,$MeV/c$^2$. Events
within $\pm2\sigma$ of the central value were retained for the
determination of the $pp\to pp\pi^0$ differential cross section.
The numbers of $\pi^0$ events deduced in this way are given for
the different energies in Table~\ref{exp:tab1}.

\begin{table}[ht]
\begin{center}
\begin{tabular}{|c|c|c|c|}                               \hline
$T_p$  & $L_{\rm{int}}$ & $N_{\pi^0}$ & $N_{\rm{bg}}/N_{\pi^0}$\\
(GeV)  & ($10^{34}$cm$^{-2}$) &             &          \\ \hline
0.508  & 0.34$\pm$0.02  & \pd131      & \pd0.094 \\ \hline
0.625  & 4.6$\pm$0.1    & 5150        & \pd0.026 \\ \hline
0.700  & 0.62$\pm$0.02  & \pd540      & \pd0.093 \\ \hline
0.800  & 6.72$\pm$0.26  & 4679        & \pd0.021  \\ \hline
1.100  & 4.08$\pm$0.16  & 1120        & \pd0.098  \\ \hline
1.400  & 7.98$\pm$0.32  & \pd779      & 0.25 \\ \hline
1.970  & 9.05$\pm$0.47  & 2065        & 0.16 \\ \hline
\end{tabular}
\caption{Summary of experimental conditions: $L_{\rm{int}}$ is the
integrated luminosity at the beam energy $T_p$ and $N_{\rm
bg}/N_{\pi^0}$ is the fractional background under the peak, where
$N_{\pi^0}$ is the number of $pp\to \{pp\}_s\pi^0$ events
registered in the angular region $\theta^{\rm cm}_{pp} <
15^\circ$.} \label{exp:tab1}
\end{center}
\end{table}

The luminosities recorded in Table~\ref{exp:tab1} were obtained by
measuring in parallel elastic proton--proton scattering using the
SP trigger. The ANKE setup detects the fast proton produced by
this reaction for cms angles between about 10$^\circ$ and
30$^\circ$. However, to avoid regions where the acceptance changes
rapidly with angle, only data from the range $15^\circ <
\theta_p^{\rm cm} < 24^\circ$ were retained for 1.4\,GeV and below
while at 1.97\,GeV the interval $17^\circ < \theta_p^{\rm cm} <
27^\circ$ was selected. After corrections for acceptance, the
numbers of events in several bins of $\theta_p^{\rm cm}$ were
compared with the values of the elastic differential cross section
taken from the SP07 solution provided by the SAID phase shift
analysis~\cite{SAID}. The precision of these predictions was
checked by looking at experimental data at small angles from which
it was assessed to be typically about $\pm4\%$, though a little
larger at 2\,GeV.

A full Monte Carlo simulation of the ANKE spectrometer has been
developed within the framework of the GEANT program~\cite{GEANT}.
This allowed us to estimate the acceptance factors for different
bins of the kinematic variables and hence, on the basis of the
luminosities given in Table~\ref{exp:tab1}, to evaluate
differential cross sections.

The experimental distributions in the final proton--proton
variables for our data at all energies are very similar to those
published at 0.8\,GeV~\cite{DymovPL}. The excitation energy
$E_{pp}$ spectrum is well described in terms of an $S$--wave $pp$
final state interaction~\cite{migdal-watson}, provided that the
Coulomb force is included. For $E_{pp}<3\,$MeV the angular
distribution of the $pp$ relative momentum vector in the final
$pp$ rest frame is consistent with isotropy, as expected for the
production of a $^{1\!}S_0$ pair.

\begin{figure}[htb]
\begin{center}
\epsfig{figure=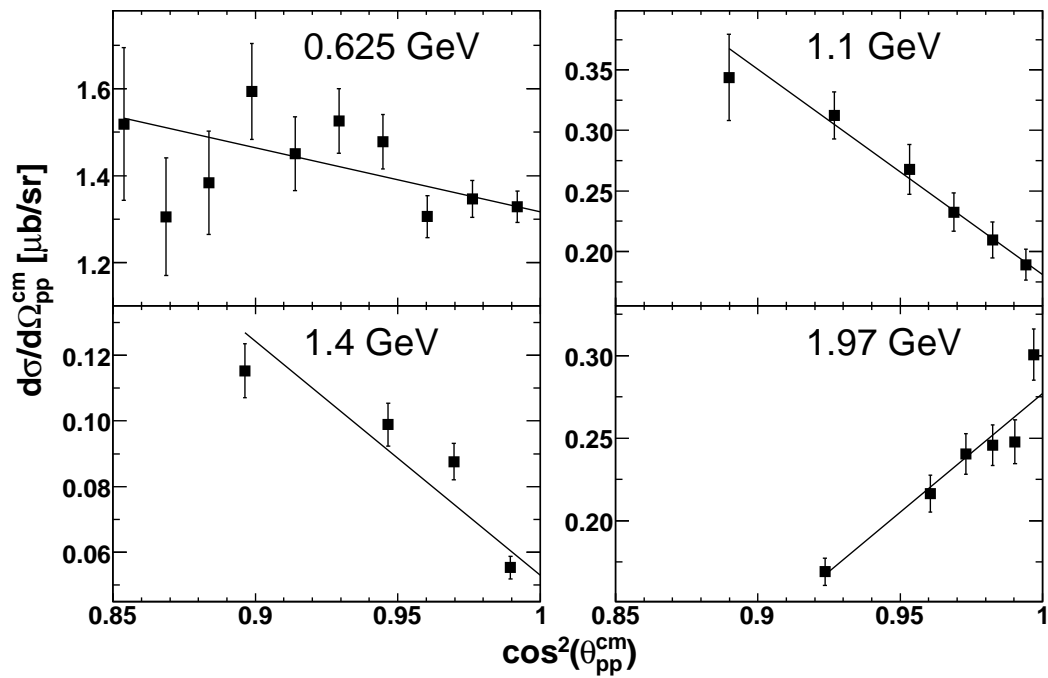, width=1\textwidth} \caption{Differential
cross section for the $pp\to \{pp\}_s\pi^0$ reaction for
proton--proton excitation energies $E_{pp} < 3\,$MeV. The results
at different beam energies are shown in terms of
$\cos^2\theta_{pp}^{\rm cm}$. The straight lines represent fits
according to Eq.~(\ref{lin-fit}) with the resulting parameters
being given for all energies in Table~\ref{res:tab-all}.}
\label{dsigf}
\end{center}
\end{figure}

Although the coverage in the diproton angle $\theta_{pp}^{\rm cm}$
with respect to the beam direction is rather limited in the ANKE
spectrometer, we already noted that the data at 0.8\,GeV showed a
strong dependence on this angle~\cite{DymovPL}. The same is true
for the other energies shown in Fig.~\ref{dsigf} where, since the
two initial protons are identical, the data are plotted as
functions of $\cos^2\theta_{pp}^{\rm cm}$. For energies of
1.4\,GeV and below, the results show a forward dip whereas at the
highest energy the cross section is maximal in the forward
direction. This is perhaps an indication that the reaction
mechanism changes with energy as different intermediate nucleon
isobars are excited.

The angular distributions have been fitted with the linear form
\begin{equation}
\label{lin-fit} \frac{\dd\sigma\pp\pp\pp\pp}{\dd\Omega_{pp}^{\rm
cm}} = \sigma_0 + \sigma_1\sin^2\theta_{pp}^{\rm cm}\,,
\end{equation}
and the values obtained for the forward differential cross section
$\sigma_0$ and the slope parameter $\sigma_1$ are recorded in
Table~\ref{res:tab-all}.

\begin{figure}[htb]
\begin{center}
\epsfig{figure=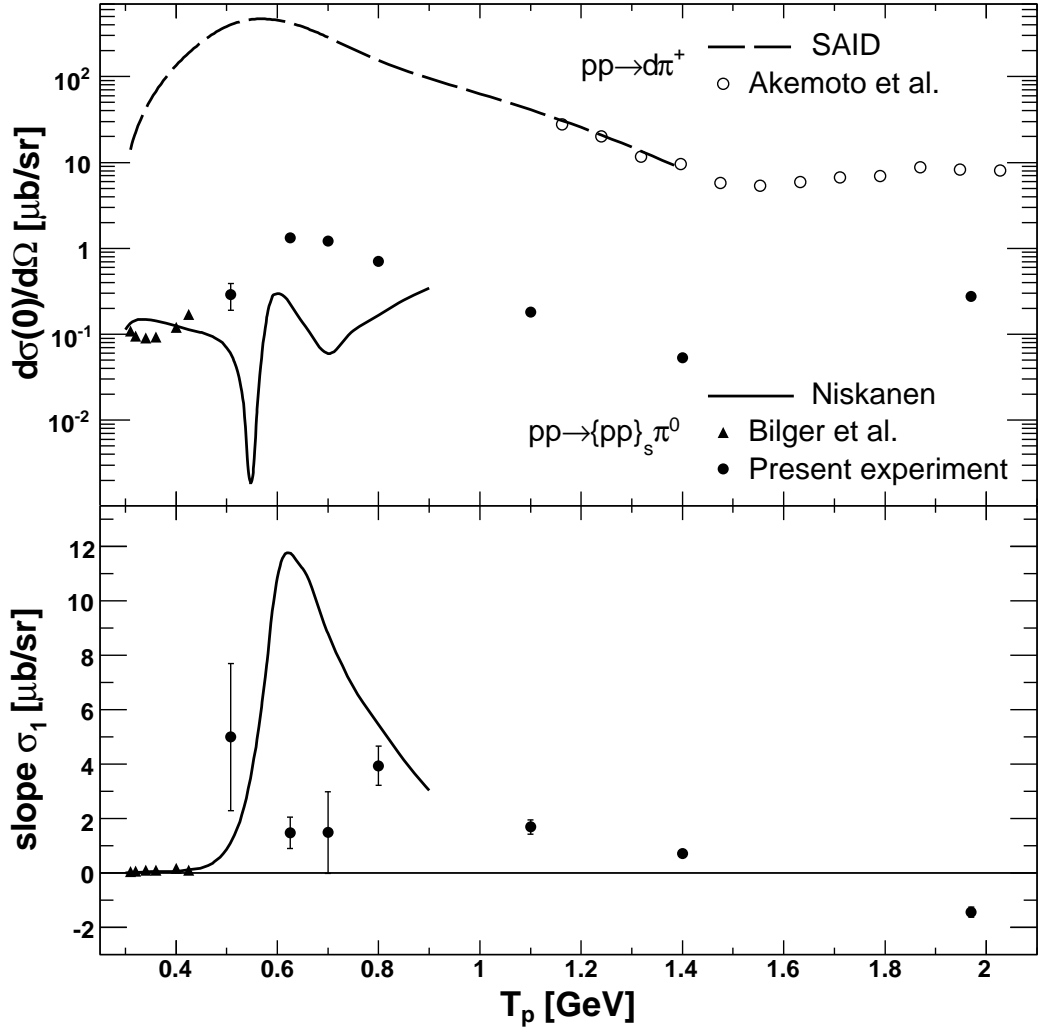, width=1\textwidth} \caption{Upper panel:
Energy dependence of the forward differential cross section for
the $pp\to \{pp\}_s\pi^0$ reaction with $E_{pp} < 3\,$MeV. The
closed circles represent the results from the present experiment
while the triangles show the low energy CELSIUS
data~\cite{BilgerNP}. For comparison we show also the
corresponding cross section for the $pp\to d\pi^+$ reaction. For
energies up to 1.4\,GeV this is represented by the dashed line
taken from the SAID parameterisation~\cite{SAID2} whereas at
higher energies the Akemoto \textit{et al.}\ data are shown as
open circles~\cite{AkemotoPL}. Lower panel: Slope parameter
$\sigma_1$ of the $pp\to \{pp\}_s\pi^0$ data, as defined by
Eq.~(\ref{lin-fit}). For both panels the solid curve corresponds
to the predictions from the $NN/\Delta(1232)N$ model of
Niskanen~\cite{Niskanen2PL}.} \label{fig-uzikov}
\end{center}
\end{figure}

Figure~\ref{fig-uzikov} summarises the results from this and a
previous experiment~\cite{BilgerNP} by showing the energy
dependence of the forward differential cross section and the slope
parameter $\sigma_1$. Although the statistical error on the
0.508\,GeV point is very large, the $pp\to \{pp\}_s\pi^0$ cross
section data suggest a significant rise from this energy to a
maximum in the 0.6--0.7\,GeV region, where $\Delta(1232)$
production is expected to be particularly strong, followed by a
monotonic decline. However, the 2\,GeV point lies much above this
trend and this, together with the slope information shown in the
lower panel, indicates that the data have entered here a different
domain. A similar tendency is seen in the $pp\to d\pi^+$ forward
cross sections, also shown on the figure, where the cross section
displays a broad minimum around 1.4\,GeV after leaving the region
influenced by the $\Delta(1232)$ isobar. However, $\pi^+$
production rises at lower energies due to contributions from the
excitation of the $\Delta(1232)N$ system in a relative $S$ wave,
which is excluded on spin--parity grounds in the $\pi^0$ case.
\vspace{3mm}

\begin{table}[ht]
\begin{center}
\begin{tabular}{|c|c|c|c|c|} \hline
$T_p$ & $\sigma_0$      & $\sigma_1$      & $\sigma_{\pi^+}$ & $R(\pi^0/\pi^+)$\\
(GeV) & ($\mu$b/sr)     & ($\mu$b/sr)     & ($\mu$b/sr)      & $\times 10^2$ \\
\hline
0.310 & $0.109\pm0.006$ & $0.0512\pm0.0016$ &   $14.1\pm0.4$   &  $8.1\pm0.5$\\
0.320 & $0.095\pm0.011$ &  $0.068\pm0.010$  &   $22.1\pm0.7$   &  $4.5\pm0.5$\\
0.340 & $0.091\pm0.023$ &  $0.111\pm0.023$  &   $41.9\pm1.3$   &  $2.3\pm0.6$\\
0.360 & $0.093\pm0.006$ &  $0.113\pm0.004$  &   $67.1\pm2.0$   & $1.44\pm0.10$\\
0.400 & $0.121\pm0.011$ &  $0.168\pm0.009$  & $\pp135\pm4\pff$ & $0.93\pm0.09$\\
0.425 & $0.171\pm0.014$ &  $0.105\pm0.011$  & $\pp189\pm6\pff$ & $0.94\pm0.08$ \\
\hline
0.508 &  $0.29\pm0.10$  &    $5.0\pm2.7$    & $\pp400\pm16\pp$ & $0.76\pm0.27$\\
0.625 &  $1.32\pm0.03$  & $\pd1.5\pm0.6\pd$ & $\pp421\pm21\pp$ & $3.26\pm0.18$\\
0.700 &  $1.21\pm0.10$  & $\pd1.5\pm1.5\pd$ & $\pp286\pm14\pp$ &  $4.4\pm0.4$\\
0.800 & $0.704\pm0.035$ & $\pd3.9\pm0.7\pd$ & $\pp155\pm8\pff$ &  $4.7\pm0.3$\\
1.100 & $0.181\pm0.013$ & $\pd1.7\pm0.3\pd$ &   $41.0\pm4\pff$ &  $4.6\pm0.5$\\
1.400 & $0.053\pm0.004$ &   $0.71\pm0.09$   & $\pd8.5\pm1.0$   &  $6.5\pm0.9$\\
1.970 & $0.277\pm0.023$ & $-1.44\pm0.19\phantom{-}$ & $\pd7.4\pm0.9$ & $39\pm6\pd$\\
\hline
\end{tabular}
\vspace{5mm} \caption{The forward differential cross section
$\sigma_0$ and slope parameter $\sigma_1$ of Eq.~(\ref{lin-fit})
for the $pp\to\{pp\}_s\pi^0$ reaction with $E_{pp} < 3\,$MeV. The
results at 0.425\,GeV and below are from Ref.~\cite{BilgerNP}.
Also shown are the forward cross sections ($\sigma_{\pi^+}$) for
$pp\to d\pi^+$~\cite{SAID2,AkemotoPL} and the ratio
$R(\pi^0/\pi^+)$ of the forward $pp\to \{pp\}_s\pi^0$ to $pp\to
\{pn\}_t\pi^+$ cross sections integrated up to 3\,MeV excitation
energy.} \label{res:tab-all}
\end{center}
\end{table}

The behaviour of the slope parameter $\sigma_1$ shown in the lower
panel of Fig.~\ref{fig-uzikov} is rather different, with a much
smoother variation and a change of sign between 1.40 and
1.97\,GeV. It is interesting to note that data on the $pp\to
d\pi^+$ differential cross section also show a small forward dip
in the 0.6--1.4\,GeV energy range~\cite{SAID2} and, furthermore,
that the sign of the forward slope changes in the 1.9--2.1\,GeV
region~\cite{AkemotoPL,AndersonPR}.

The only theoretical estimates of the cross section and slope
parameter for the $pp\to \{pp\}_s\pi^0$ reaction for the small
$E_{pp}$ kinematical domain have been made by
Niskanen~\cite{Niskanen2PL}. The cross section predictions shown
in Fig.~\ref{fig-uzikov}, which reflect the delicate interferences
between contributions from $NN$ and $\Delta(1232)N$ intermediate
states, do illustrate a displacement of the $\Delta(1232)N$ peak
upwards by about 100\,MeV compared to $pp\to d\pi^+$. Although
this agrees with our observations,  Niskanen's predicted minimum
around 0.7\,GeV is in complete contradiction to our findings of a
maximum in this region. Furthermore we see no sign of any big peak
in the slope $\sigma_1$ around 0.6\,GeV so that any agreement with
our 0.8\,GeV data seems fortuitous.

To facilitate the comparison of $\pi^+$ and $\pi^0$ production,
the values of both the $pp\to\{pp\}_s\pi^0$ and $pp\to d\pi^+$
forward cross sections are given in Table~\ref{res:tab-all}.
However, to compare the strengths of pion production leading to
the $S$--wave spin--triplet and singlet $NN$ states, we require
rather data on the $pp\to \{pn\}_t\pi^+$ channel integrated up to
3\,MeV excitation energy. Using final--state--interaction theory,
this can be approximated for low $pn$ excitation energies by the
$pp\to d\pi^+$ data through~\cite{FaeldtWilkinPL}
\begin{eqnarray}
\nonumber \frac{\dd\sigma}{\dd \Omega} (pp \rightarrow
\{pn\}_t\pi^{+}) &\approx& \frac{\dd\sigma}{\dd\Omega}
(pp\rightarrow d\pi^{+}) \times \int_0^{k_{\rm max}}\frac{k^2}{
\pi\alpha_t(k^2+\alpha_t^2)}\,\dd k \\
&\approx& 0.096\times\frac{\dd\sigma}{\dd\Omega} (pp\rightarrow
d\pi^{+})\,,
\end{eqnarray}
where $k$ is the relative momentum in the final $NN$ system,
$k_{\rm max}^2/m_N = 3\,$MeV, and $\alpha_t^2=m_NB$, with $B$
being the deuteron binding energy. Though there are some
deviations from this, due in part to the $pn$ tensor force, the
approximation gives a plausible description of experimental
data~\cite{Abdel-BaryPL}.

\begin{figure}[htb]
\begin{center}
\epsfig{figure=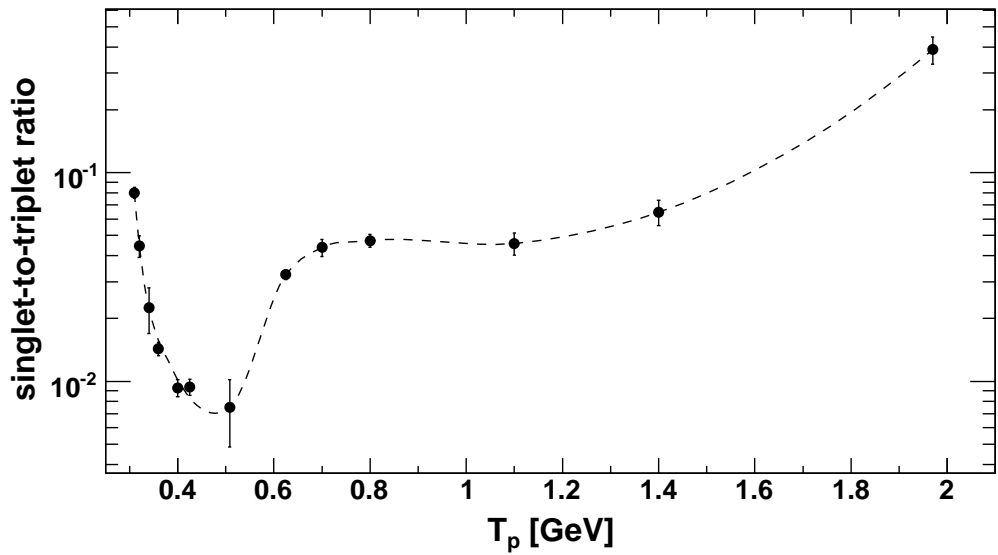, width=1\textwidth} \caption{Energy
dependence of the ratio $R(\pi^0/\pi^+)$ of the forward $pp\to
\{pp\}_s\pi^0$ and $pp\to \{pn\}_t\pi^+$ cross sections integrated
up to 3\,MeV excitation energy. The curve is drawn to guide the
eye.} \label{fig-zeta}
\end{center}
\end{figure}

A common feature of the $R(\pi^0/\pi^+)$ ratio seen in
Fig.~\ref{fig-zeta} is its relatively low value. Below 1.4\,GeV it
is typically a few per cent, which is significantly smaller than
the trivial spin--statistics factor of 1/3. These conclusions are
not changed significantly if the difference between the
spin--singlet and triplet final state interactions are taken into
account~\cite{UzikovEPJ}.

The dip in the ratio in the 0.5\,GeV region is in part a
reflection of the fact that the $S$--wave $\Delta(1232)N$
intermediate state is forbidden for $\pi^0$ production whereas it
plays a vital role in the case of the $\pi^+$. However it is not
at all clear why $R(\pi^0/\pi^+)$ is small in the 1\,GeV region
where the contribution from the $\Delta(1232)$ maximum is much
reduced. On the other hand, the much larger value at 1.97\,GeV is
yet another indication that the reaction mechanisms might be
different at higher energies.

In summary, we have measured the differential cross section for
the $pp\to\{pp\}_s\pi^0$ reaction at seven beam energies from 0.5
to 1.97\,GeV under the specific kinematic conditions where the
proton--proton excitation energy is below 3\,MeV and the cms angle
between the diproton momentum and the beam axis is less than
$18^0$. The observed form of the $E_{pp}$ spectra and isotropy of
the angular distribution in the diproton cms are consistent with
the assumption that the two final protons are in the $^{1\!}S_0$
state. Except for the highest energy, the data all show a sharp
minimum in the forward direction and the ratio there of the
differential cross sections for the $pp\to\{pp\}_s\pi^0$ and
$pp\to d\pi^+$ reactions is below 1\%. The situation changes
radically  at 2\,GeV when a forward maximum is observed with a
much enhanced value of $R(\pi^0/\pi^+)$. This must be a reflection
of a different reaction dynamics at high energies. A broad peak
around 0.6--0.7\,GeV is observed in the energy dependence of the
zero--degree differential cross section. The data are consistent
with the predicted displacement of the $\Delta(1232)$ maximum to
higher energies for $\pi^0$ production but the other features of
our results are not explained by a microscopic
model~\cite{Niskanen2PL}. This could, for example, be due to phase
differences between intermediate $NN$ and $\Delta(1232)N$
contributions and much theoretical work still remains to be done.

We have attempted to study the relative amplitudes for the
production of spin--singlet and spin--triplet final $NN$ states by
using a simplistic model that links the $pp\to d\pi^+$ and $pp\to
\{pn\}_t\pi^+$ cross sections. The resulting $R(\pi^0/\pi^+)$
ratio shows that singlet production in the forward direction
remains small at energies even above the $\Delta(1232)$ excitation
region.

Further data are clearly required in the 0.5--0.6\,GeV region
where the Niskanen model should be more reliable than at higher
energies to see if there is indeed any trace of his predicted
structure there. The transition region above 1.4\,GeV is also of
interest. Does the spin--singlet suppression continue when higher
isobars enter?

For a $^{1\!}S_0$ final state there are only two spin--amplitudes
for the $pp\to\{pp\}_s\pi^0$ reaction. Since we observe a very
strong angular variation over our acceptance region, it is then
likely that the analysing power will also be significant there. It
is therefore planned that in the near future measurements of $A_y$
will be carried out at COSY~\cite{Kulikov} and eventually
spin--correlation studies might be achievable~\cite{SPIN}. This
ensemble of data should provide a significant challenge for theory
in this, one of the simplest pion--production reactions.

\noindent
\textbf{Acknowledgments}

This work was supported in part by the BMBF grant ANKE COSY--JINR
and COSY--FFE. We are grateful to other members of the ANKE
Collaboration for their help with this experiment and to the COSY
crew for providing such good working conditions. J.A.~Niskanen
kindly provided the numerical values of the calculations of
Ref.~\cite{Niskanen2PL}.

%
%

\end{document}